\documentclass[twocolumn,prl,showpacs,10pt]{revtex4}
\usepackage{graphicx}
\usepackage{epsfig}

\newcommand{\lsim}{\,\raise 0.4ex\hbox{$<$}\kern -0.8em\lower 0.62ex\hbox{$\sim$}\,}
\newcommand{\gsim}{\,\raise 0.4ex\hbox{$>$}\kern -0.7em\lower 0.62ex\hbox{$\sim$}\,}

\def\be{\begin{equation}}
\def\ee{\end{equation}}
\def\bea{\begin{eqnarray}}
\def\eea{\end{eqnarray}}
\newcommand\de{\mathrm{DE}}

\begin{document}

\title{The dark degeneracy: On the number and nature of dark components}
\date{February 27, 2007}

\author{Martin Kunz}
\email{Martin.Kunz@physics.unige.ch}
\affiliation{D\'epartement de
Physique Th\'eorique, Universit\'e de
Gen\`eve, 24 quai Ernest Ansermet, CH--1211 Gen\`eve 4, Switzerland}

\begin{abstract}
We use that gravity probes only the total energy momentum tensor
to show how this leads to a degeneracy for generalised dark energy
models. Because of this degeneracy, $\Omega_m$ cannot be measured.
We demonstrate this explicitely by showing that the CMB and supernova
data is compatible with very large and very small values of $\Omega_m$
for a specific family of dark energy models.
We also show that for the same reason interacting dark energy is always
equivalent to a family of non-interacting models. We argue that it is
better to face this degeneracy and to parametrise the actual observables.
\end{abstract}


\maketitle


{\bf \em Introduction~~} Over the last decades, cosmology has turned into
an experimental science, with more and more high-quality data
becoming available. The most surprising conclusion from this
data is the existence of a dark contribution to the energy
density in the universe, which interacts through gravity with
normal matter, and which seems to make up 95\% of the total
energy density today. To understand the nature of this dark
sector is correspondingly considered one of the most important 
tasks in cosmology.

The line element of a perfectly homogeneous and isotropic space with
zero curvature is
\be
ds^2 = -dt^2 + a(t)^2 dx^2
\ee
with only one degree of freedom, the scale factor $a(t)$. The
energy-momentum tensor has to be compatible with perfect
homogeneity and isotropy, which means that it has to have the
form of a perfect fluid,
\be
T_\mu^\nu = \mathrm{diag} (-\rho(t),p(t),p(t),p(t)) .
\ee
There are two degrees of freedom in the energy momentum
tensor (EMT), $\rho(t)$ and $p(t)$. The Friedmann equations can
only fix the behaviour of one of them, conventionally taken
to be $\rho(t)$, while the other one is an a-priori free
function of time, describing the physical properties of the
perfect fluid. In cosmology one usually poses $p(t)=w(t)\rho(t)$
so that $w(t)$ is now a free function.

Photons and baryonic matter are detected through their
non-gravitational interactions, and their contribution
to $T_{\mu\nu}$ can be measured directly. But the dark
sector, by definition, is only constrained through gravity,
which depends only on the {\em total} energy momentum tensor. Gravity
therefore only constrains the total $w(z)$. Any further
freedom, like sub-dividing the dark EMT into dark matter
and dark energy, or introducing couplings between the dark
constituents, cannot be directly measured and will introduce
degeneracies.


{\bf \em Background constraints on $\Omega_m$ and $w(z)$~~}
One generally postulates that the ``energy excess'' in the universe,
the dark matter and dark energy, are two different components, with
the dark matter being characterised by $w_m=0$ and a relative energy 
density today of $\Omega_m=8\pi G \rho_m(t_0)/(3H_0^2)$.

However, nothing stops us from adding the energy-momentum tensor
of the dark energy and of the dark matter together into a combined
``dark fluid'' EMT. This provides also a solution
of the Friedmann equations, but with a different equation of state.
Worse, we can just as well split that dark fluid EMT
{\em arbitrarily} into one part with $w=0$ and another part with a time-varying
equation of state. To show this explicitely, we notice that 
for a flat universe composed of matter and dark energy with unknown
$w(z)$, and given $H(z)$, we find from the Friedmann equations
(see e.g.~\cite{bdk})
\be
w(z) = \frac{H(z)^2-\frac{2}{3} H(z) H'(z) 
(1+z)}{H_0^2 \Omega_m (1+z)^3-H(z)^2}
\label{eq:wh}
\ee
where we used the observationally more relevant redshift $z$ as
time variable, and $H'=dH/dz$.
This means that for {\em any} choice of $\Omega_m$ we find a $w(z)$
which reproduces the measured expansion history of the universe.
In other words, $\Omega_m$ cannot be measured if $w(z)$ is not
known.

This fundamental point does not change if we add radiation and
baryons. Their abundance can be measured in different ways
thanks to their interactions. Curvature is also a dark component
and exhibits a similar degeneracy \cite{chiba,bruce}, but it can at
least in theory  be distinguished from the other dark
components due to its different geometric structure \cite{curv,GB}.
We will therefore limit this discussion to flat space.

\begin{figure}[ht]
\epsfig{figure=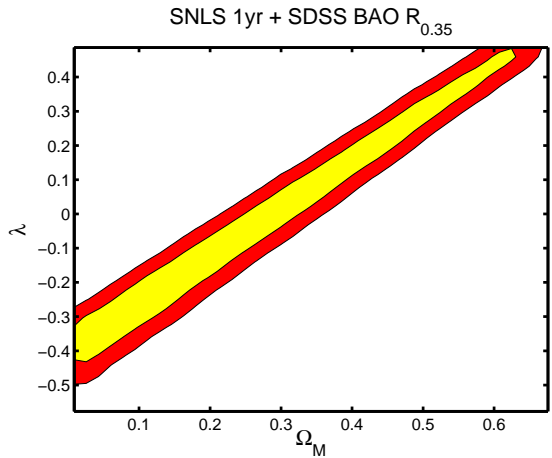,width=3.2in}
\caption{We try to determine $\Omega_m$ with the SNLS 1yr data
and the $R_{0.35}$ constraint from the baryon acoustic oscillations
measured by SDSS LRG,
but using an equation of state that exhibits a degeneracy between
dark matter and dark energy.
We find that the background data cannot determine $\Omega_m$.
}
\label{fig:snls1}
\end{figure}

To illustrate the real nature of this fundamental degeneracy we
first notice that for a flat $\Lambda$CDM concordance-like model
with $w=-1$ and a given $\Omega_m$ the family of
dark energies with the same $H(z)$ lead to the following family
of equation of state parameters as a function of the apparent
dark matter density $\hat{\Omega}_m$:
\be
\hat{w}(z) = - \frac{1-\Omega_m}{(1-\Omega_m) +(\Omega_m-\hat{\Omega}_m) (1+z)^3} .
\label{eq:w_lcdm}
\ee
Inspired from this result we set
$w(z) = - 1/(1-\lambda (1+z)^3)$ and try to measure
simultaneously $\lambda$ and $\Omega_m$ from the
SNLS supernova data \cite{snls} and the $R_{0.35}$ constraint 
from the baryon acoustic oscillations (BAO)
measured by SDSS LRG \cite{sdss_bao}. As expected, we find a strong degeneracy
and no limit on $\Omega_m$, see Fig.~\ref{fig:snls1}. 
This does not change if we add further background data,
as this is a fundamental degeneracy
that is present in {\em all} probes of the expansion history.

For our choice of $H(z)$ we find, not surprisingly, that 
$\rho_\de(z)\propto 1-\lambda (1+z)^3$. 
In the range $\lambda<0$ the dark energy absorbs
part of the dark matter, and becomes more similar to it. For
$\lambda>0$ the dark energy has to evolve in the different
direction, becoming a kind of ``anti dark matter''.
In order to achieve this, $\rho_\de$ becomes negative. This
may appear strange, but is very similar to the behaviour of
a scalar field in a potential which is negative for some
field values. Apparently negative energy-densities also appear in some
theories of modified gravity with a non-standard Friedmann
equation \cite{odin,fR}. 
Since we have derived the form of $w(z)$
from a well-behaved $H(z)$ we know that even an apparently strange
equation of state leads to a well-behaved expansion
history of the universe.

Measuring only the expansion history of the universe therefore
does not allow us to make separate statements about the dark
matter and the dark energy. Rather, $\Omega_m$ becomes a parameter
which enumerates a family of dark energy models according to
Eq.~(\ref{eq:w_lcdm}). All members of this family lead to exactly
the same expansion history. From an experimental point of view,
they should be regarded as forming an equivalence class of models. 
It is possible that the all matter is baryonic, and that the
dark energy is characterised by $w(z)=-1/(1+0.3(1+z)^3)$. Analysing
this scenario with the usual parametrisations of the equation of
state, we would be led to conclude wrongly that $\Omega_m\approx0.3$
and that the dark energy has a fine-tuning problem. 

If we assume that there is a period of matter domination at 
high redshift so that $H(z)\propto(1+z)^{3/2}$, then the numerator
of Eq.~(\ref{eq:wh}) is zero and so the behaviour of the dark
energy approaches that of matter at high redshift. This can only
be avoided if the denominator vanishes as well, which singles
out one specific value of $\Omega_m$ for which $w$ does not go
to zero at high redshift. It is this value which is conventionally 
considered to be the ``true'' one, but we have to be aware that this
is a philosophical choice which cannot be supported by
experimental evidence. Furthermore, this is the choice which,
by construction, creates a fine-tuning problem for the dark energy, as
its relative density will decrease with $z$.

Given this degeneracy, we can still compare models and exclude
those which agree significantly worse with data than others. 
However, we cannot actually measure quantities like $\Omega_m$
and $w(z)$ with background data alone.

It is also worrying that this degeneracy seems to have escaped
notice of the numerous analyses trying to constrain $w(z)$ with
supernova data and other distance measures. This illustrates
once more that parametrisations impose strong priors on the kind
of dark energy models probed, as e.g. argued in \cite{bck}. 


{\bf \em Interacting dark energy models~~}
The above effect has also implications for other models, for
example those where the dark energy and the dark matter are
interacting. In this case their total energy momentum tensor has
to be conserved for consistency with the Bianchi identities of
the Einstein tensor (e.g.~\cite{am99}),
\be
\left(T_{\mu\nu}^{(m)}+ T_{\mu\nu}^{(\de)}\right)_{;\mu} = 0 .
\label{eq:totcons}
\ee
But this means that the total EMT is just of the same class
as the one discussed in the previous section. We can either keep it as a
single unified dark fluid model, we can divide
it into a coupled dark matter -- dark energy system, or we can
also divide it into uncoupled dark matter and dark
energy. The only quantity that we measure is again $H(z)$ and
this only fixes the {\em total} EMT and does not tell us
anything about how it should be split. Correspondingly the
interaction between dark matter and dark energy is also perfectly
degenerate with the dark energy $w(z)$. In other words, we cannot
measure it unless we fix {\em both} $\Omega_m$ and $w(z)$.

What happens is the following: The interaction modifies the conservation
equations of the two dark components,
\bea
\dot{\rho}_m + 3 H \rho_m &=& C(t) \\
\dot{\rho}_{\de} + 3 H (1+w) \rho_{\de} &=& -C(t)
\eea
The sum of the two equations has to be zero in order to satisfy
Eq.~(\ref{eq:totcons}). These two equations, together with one
of the Friedmann equations, determine $H(z)$. From $H(z)$ we
can then derive a family of uncoupled models, using
Eq.~(\ref{eq:wh}), or also families of models with other interactions.

Let us look in more detail at an especially simple case that
is often used, e.g. 
\cite{MaSaAm}, where
$C = \gamma H \rho_m$ and $\gamma$ is constant, and we also
take $w$ to be constant. 
This makes it easy to integrate the above equations, giving
\bea
\rho_m &=& \rho_m^{(0)} (1+z)^{3-\gamma} \\
\rho_\de &=& \left(\rho_\de^{(0)} + \rho_m^{(0)} \frac{\gamma}{\gamma + 3 w}\right)
(1+z)^{3(1+w)} \nonumber\\ 
&&- \rho_m^{(0)} \frac{\gamma}{\gamma + 3 w} (1+z)^{3-\gamma}
\eea
and the Hubble parameter is (assuming flatness again)
\bea
H^2 &=& H_0^2 \left[ \Omega_m \left(1-\frac{\gamma}{\gamma + 3 w}\right) (1+z)^{3-\gamma} \right. \nonumber \\
 && \left.+ \left( 1- \frac{3 \Omega_m w}{\gamma + 3 w}\right) (1+z)^{3(1+w)}  \right] .
\eea
In the case where our data is actually due to a decaying cosmological
constant ($w=-1$) with $\Omega_m=0.3$ and a constant $\gamma$, we find 
that we can just as well fit
it with uncoupled dark matter and dark energy with an equation of state
\be
\hat{w}(z) = \frac{-0.3 \gamma + (\gamma-2.1)(1+z)^{-3+\gamma}}
{0.9 + (1+z)^\gamma \left(\hat{\Omega}_m(\gamma-3) - (\gamma-2.1)/(1+z)^3 \right)}
\label{eq:w_g}
\ee
We are again free to choose an apparent $\hat{\Omega}_m$ different from $0.3$.
Conversely, given a non-interacting dark matter -- dark energy model
with a certain $w$, we can pretend that we are actually dealing with a
coupled cosmological constant by solving Eq.~(\ref{eq:w_g}) for $\gamma$.
(Although one would have to generalise the above discussion to time-varying
couplings in order to do that.) As has been noticed before \cite{wandelt,coup_phant}
this could be used to replace phantom dark energy (with $w<-1$) with a
non-phantom interacting model.

It is only possible to constrain the couplings by imposing a prior
on the space of possible models. It is important that we are aware
of this limitation, as we do not know the nature of the dark energy.
We can always trade off
a specific form of the interaction against a different $w(z)$ and a
change in $\Omega_m$. 
Finally, coupling dark matter and dark energy does
not lead to any new phenomena in the dark sector, beyond those which
can be achieved by general uncoupled dark energy.


{\bf \em Beyond the background~~}
Is this degeneracy just a problem at the background level, and can it
be broken when we take into account that the universe is not homogeneous
and isotropic? In general, it cannot: the fundamental reason of this
``dark degeneracy'' stems directly from the structure of the full
Einstein equations,
\be
G_{\mu\nu} = 8 \pi G T_{\mu\nu} .
\ee
The Einstein tensor $G_{\mu\nu}$ on the left hand side describes
the geometric aspects of General Relativity, while the energy-momentum
tensor on the right hand side defines the energy and pressure content.
Although the equations are highly non-linear in $g_{\mu\nu}$, they
are linear in the components of $T_{\mu\nu}$. If the only information
on a part of $T_{\mu\nu}$ comes from gravitational probes, as is
true by definition for ``dark stuff'', then we can decompose this
part in any way we want -- we cannot learn anything about the
sub-parts, only about the whole.

As an illustration, in first order perturbation theory the dark fluids influence the ``bright
side'' through the gravitational potentials $\phi$ and $\psi$ which describe
the scalar metric perturbations. The physical properties of fluids are described
by two additional quantities, for example the pressure perturbations
$\delta p$ and the anisotropic stresses $\Pi$. They can be different
for each fluid, but as discussed e.g.~in \cite{mawa,phantom} it is always
possible to combine several fluids into a single one with an effective
$\delta p$ and $\Pi$. This single fluid then contributes in exactly the
same way to $\phi$ and $\psi$ as its constituent fluids. Conversely, 
any fluid can be split in a basically arbitrary way into sub-fluids.

\begin{figure}[ht]
\epsfig{figure=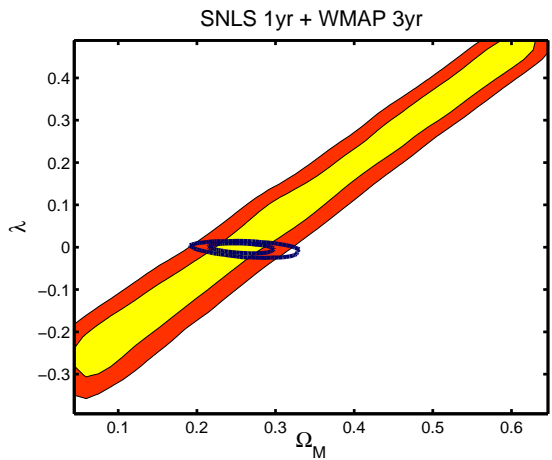,width=3.2in}
\caption{This figures shows that supernova and CMB data together
cannot measure $\Omega_m$ for generalised dark energy models. The
filled contours show $1$ and $2\sigma$ limits for a model with 
$c_s^2 =0$, while the open contours show the limits for an
effective scalar field model. The lower limit on $\Omega_m$ is
due to the baryons which we know to exist.}
\label{fig:cluster}
\end{figure}

We can escape the degeneracy by considering
specific models, for example scalar
field dark energy for which we know that $\delta p$ is given by a 
rest-frame sound speed $c_s^2=1$, and $\Pi=0$. In this way we basically define the
dark energy to be the dark part which does not cluster. This may be
a reasonable way to break the degeneracy, but we should not forget
that it may well be that there is only one dark fluid that clusters
partially, rather than one strongly clustering and one non-clustering
part. Also modifications of gravity like DGP
can act effectively like a clustering form of dark energy \cite{dgp}.
We show in Fig.~\ref{fig:cluster} that indeed non-clustering
dark energy leads to a well-defined $\Omega_m$ when using the WMAP3
CMB data \cite{wmap3} together with the SNLS 1 yr supernova data
(open contours) and using the equation of state parameter (\ref{eq:w_lcdm}).

The general expression for $\delta p$ along the degeneracy can be quite
complicated. In our example the dark matter has $w=c_s^2=0$ and
the cosmological constant does not carry perturbations so that
we find $\delta p = 0$ in the Newtonian gauge. Even 
just setting $c_s^2 = 0$ suffices to restore the degeneracy 
(filled contours of Fig.~\ref{fig:cluster}). Does dark
energy cluster? We do not know. Fig.~\ref{fig:cluster} shows that
clustering dark energy is perfectly compatible with CMB and
supernova data. These results, derived
using a modified version of CAMB \cite{camb} also illustrate
the dangers of using standard parametrisations of experimental results
for studying non-standard dark energy models. The shift parameter $R$
of the CMB as well as $A$ parameter of BAO data contain $\Omega_m$
explicitely. They are therefore only valid for very specific models.

As argued above using the full Einstein equation, this game can be
played to all orders. For example, galaxy rotation curves fix the
amount of clustered dark stuff. So we can use this to determine
the amount of dark matter only if we (arbitrarily)
impose that dark energy does not cluster. Stars feel gravity,
not the dark matter itself.


{\bf \em Conclusions~~}
Gravity only responds to the total energy momentum tensor $T_{\mu\nu}$.
We have been able to measure the physical properties of some of its
constituents, like baryons and photons, in laboratory experiments, and 
their contribution to $T_{\mu\nu}$ can be separated out. However,
by definition the dark parts do not show up in laboratory experiments
and interact only with gravity, so that we can only probe the total
dark EMT. With probes of the background evolution, 
we can measure for example $H(z)$, corresponding to the overall
dark equation of state parameter $w_\mathrm{tot}(z)$. At the level of first order
perturbation theory, the observables are for example the gravitational
potentials $\phi$ and $\psi$, or the overall anisotropic stress $\Pi$ 
and the pressure perturbations $\delta p$ of the dark sector.

Conventionally, the dark sector is sub-divided into dark matter and
dark energy. 
Here we have shown explicitely that, being in a state of total ignorance 
about the nature of a single one of the dark
components, we can also not completely measure the others.
In this situation the separation into dark matter and
dark energy becomes merely a convenient parametrisation without
experimental reality. Indeed, we need to impose a specific
condition to make this split well-defined, for example that
the dark energy has to vanish at high redshift, or that the
dark energy constitutes the non-clustering part of the dark
EMT. Such a split is useful for {\em testing} specific
models of the dark constituents, but gravity alone cannot 
{\em measure} their physical properties independently.

We have also demonstrated the dangers of using fitting formulae for 
experimental results, like the peak locations
in the CMB or the baryonic oscillations.
 Great care has to be
taken when analysing non-standard dark energy models with such formulae, and it
is generally preferable to err on the side of caution and to recalculate
the actually measured experimental quantities ab-inito.

We now need either a theoretical breakthrough
which produces a well-motivated model of all things dark and of their
properties, which agrees with the data, or else a direct 
(non-gravitational) detection of the
dark matter, for example with the LHC at CERN or with scattering
experiments. Fixing the abundance and evolution of the dark
matter (and putting strong constraints on its 
couplings to other constituents) allows us to remove the word ``dark'' from its name
and subsequently to probe the remaining part of $T_{\mu\nu}$ which
describes the dark energy -- assuming that such a split is realised
in nature, which we do not know yet.
Dark matter experiments are therefore of the highest
importance for dark energy studies as well.
Without their results, we can only deduce the {\em overall}
properties of the dark side from cosmological data and state
that they are compatible (or not) with a given model (e.g.
$\Lambda$CDM).

\begin{acknowledgments}
It is a pleasure to thank
Luca Amendola, Ruth Durrer and Domenico Sapone for numerous
interesting discussions. MK acknowledges
funding by the Swiss NSF.
\end{acknowledgments}

\end{document}